\documentclass[showpacs,aps,twocolumn]{revtex4}
\usepackage{amsmath}
\usepackage{amsfonts}
\usepackage{graphicx}

\newcommand{\p}{\prime}

\newcommand{\cre}[2]{\ensuremath{#1_{#2}^{\dagger}}}
\newcommand{\ann}[2]{\ensuremath{#1_{#2}^{}}}
\newcommand{\cres}[2]{\ensuremath{#1_{#2\sigma}^{\dagger}}}
\newcommand{\anns}[2]{\ensuremath{#1_{#2\sigma}^{ }}}
\newcommand{\crebs}[2]{\ensuremath{#1_{#2\bar{\sigma}}^{\dagger}}}
\newcommand{\annbs}[2]{\ensuremath{#1_{#2\bar{\sigma}}^{{ }}}}
\newcommand{\aver}[1]{\langle #1 \rangle}
\newcommand{\acom}[1]{\{#1\}}

\begin{document}


\title{Negative differential resistance due to the resonance coupling of 
a quantum-dot dimer}

\author{S. D. Wang}
\email{wangsd@ust.hk}
\author{Z. Z. Sun}
\author{N. Cue}
\author{X. R. Wang}
\affiliation{Physics Department, The Hong Kong University of
Science and Technology, Clear Water Bay, Hong Kong SAR, China}

\date{\today}

\begin{abstract}
  Electron tunneling through a coupled quantum-dot dimer under a
  dc-bias is investigated. We find that a peak in the $I$-$V$ curve
  appears at low temperature when two discrete electronic states in
  the two quantum dots are aligned with each other -- resonance
  coupling. This leads to a negative differential resistance. The peak
  height and width depend on the dot-dot coupling. At high
  temperature, the peak disappears due to thermal smearing effects.
\end{abstract}



\maketitle


\section{Introduction}

A large number of studies have focused on nanostructure materials,
such as quantum dots (QDs)\cite{Kastner}, as we move into the era of
nanoscience and nanotechnology. This is due largely to academic
interests and their potential applications. It has been proposed that quantum
dots are used as building blocks of electric circuits\cite{device}
and even of quantum logic gates\cite{qgates}. Since most of these
building blocks involve many QDs, many experimental\cite
{CB-e,Vaart,mole} and theoretical\cite{CB-t,Fong,Palacios} 
investigations are focused on more than one QD system, especially 
two-QD system -- QD dimer. In a two QD system, not only the charging 
effect but also the interdot coupling and the alignment of electronic 
states in the two QDs play important roles. Because of the interdot
coupling, electrons can be shared by both QDs. A state similar to the
covalent state in a molecule can be formed in a QD dimer. This state
can be manipulated by varying the external parameters of the system,
such as the interdot coupling\cite{mole}. Therefore QD dimers are
proposed to be candidates for building quantum logic gates\cite{qgates}.
The interdot coupling also yields new features in the Coulomb
Blockade conductance spectroscopy. For example, a system
with two isolated identical QDs has an oscillation structure in the
conductance spectroscopy in the Coulomb blockade regime. By tuning
the interdot coupling, those peaks are split into two peaks
each\cite{CB-e, CB-t}.  The alignment of electronic states in two QDs
makes the $I$-$V$ characteristics of a dimer with two QDs in series
much different from that of a single QD system with a step-like
structure. The $I$-$V$ curve of a QD dimer has many peaks and these
peaks are due to the resonant tunneling when two electronic states in
the two QDs are aligned\cite{Vaart, Fong}.

Most of the previous studies are on a QD dimer with two QDs in series. 
In this paper, we shall study a system with two QDs coupled in parallel 
with source and drain leads. One QD (QD1) is connected to both source 
and drain leads while the other QD (QD2) is connected to only one lead. 
This configuration may arise when one performs STM experiments on 
quantum dots on a substrate. We show that the alignment of electronic 
states in the coupled QD dimer, which we shall call resonance coupling, 
can modify greatly the electron transport. 
We find that a peak in $I$-$V$ curve occurs when two electronic states
in the two QDs are aligned, leading to a negative differential
resistance (NDR) at low temperature. In the presence of the 
electron-electron (e-e) interaction, the peak splits into two. 

The present paper is organized as follows. Our model and the method used to 
calculate the $I$-$V$ curves are described in Sec.~\ref{sec:model}.
Then, we present, in Sec.~\ref{sec:results}, our calculated results 
for different cases. The numerical evidence of a NDR in our model is 
given. We shall also provide the explanation of the occurrence 
of the NDR. The summary in
Sec.~\ref{sec:summary} concludes the presentation.


\section{Model and method}
\label{sec:model}
 
\begin{figure}[htb]
  \begin{center}
    \includegraphics[width=6.5cm,height=4.5cm]{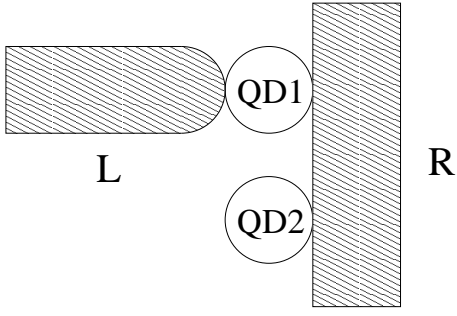}
  \end{center} 
  \caption{Schematic diagram of a system with two coupled quantum dots
    connected to source and drain leads in parallel. One quantum dot
    (QD1) is connected to both the left (L) and the right (R) external
    metallic leads. The other quantum dot (QD2), is only connected to
    R-lead.} 
  \label{fig:system}
\end{figure}

As shown in Fig.~\ref{fig:system}, we consider a QD dimer with two
coupled QDs connected to metallic leads in parallel. One of the leads 
act as a source, say the left lead (L-lead) in the figure, and the 
other (the right lead) as a drain. One quantum dot (QD1) is connected 
to both external leads while the other quantum dot (QD2) is only 
connected to one of these leads, say the right lead (R-lead) as shown 
in the figure. Thus, electrons, flowing from the L-lead to the R-lead 
through QD2 must tunnel to QD1 first. This set-up can describe an STM 
tunneling experiment of a QD (QD1) on a substrate while there is an 
adjacent QD (QD2) coupled to QD1.

We shall model both the L- and R-leads as one-dimensional (1D) 
non-interacting Fermi gases. We assume that the electron level 
spacing in both QDs is large in comparison with the level 
broadening due to the tunneling process and the thermal effect. Thus, 
the spectra of both QDs are discrete. For simplicity, we consider 
that there is only one energy level with spin degeneracy in each QD. 
The electron-electron (e-e) interaction in a mesoscopic system is 
usually larger than or comparable to other energy scales like  
the typical level spacing and the thermal energy. Hence the Coulomb
blockade effect plays an important role in electron transport when QDs
are weakly coupled to external leads. We shall consider the e-e
interaction in our model. And we shall describe the tunneling process
between leads and QDs with a tunneling Hamiltonian.

The total Hamiltonian of this system can be expressed as 
\begin{equation}
  H = H_{L,R} + H_c + H_T,
\end{equation}
where $H_{L,R}$ is the Hamiltonian of two 1D ideal metallic leads,
$H_c$ is the Hamiltonian of the central region with two coupled
QDs, and $H_T$ is the tunneling Hamiltonian.

The Hamiltonian of two 1D ideal metallic leads is
\begin{equation}
  H_{L,R} = \sum_{l,k,\sigma} \epsilon_{l,k}
  \cres{c}{l,k}\anns{c}{l,k},
\end{equation}
where $\sigma=\uparrow,\downarrow$ is the spin index, and
$\cres{c}{l,k}$ ($\anns{c}{l,k}$) is the creation (annihilation)
operator of an electron with energy $\epsilon_{l,k}$ and spin $\sigma$
in lead $l=L,R$, the left or right leads.

The Hamiltonian of the central region is
\begin{align}
  H_c &= \sum_{i=1,2 \atop \sigma} \epsilon_i \cres{d}{i}\anns{d}{i} +
    \sum_{\sigma} \bigl(t\cres{d}{1}\anns{d}{2} 
  + \mathrm{H.c.} \bigr) \nonumber \\
  &\qquad\qquad +\sum_{i=1,2} U_i
    \ann{n}{i\uparrow}\ann{n}{i\downarrow}, 
\end{align}
where $\cres{d}{i} (\anns{d}{i})$ is the creation (annihilation)
operator of an electron with energy $\epsilon_i$ and spin $\sigma$ in
QD $i$.  $t$ is the hopping energy between the states
$\epsilon_1$ and $\epsilon_2$ in the two QDs. The terms containing
$U_i$ describe the e-e interactions in QD $i$.

The tunneling Hamiltonian is given as
\begin{equation}
  H_T = \sum_{l,k,\sigma \atop i=1,2}
  V_{l,k\sigma,i}\cres{c}{l,k}\anns{d}{i} + \mathrm{H.c.},
\end{equation}
where $V_{l,k\sigma,i}$ is the coupling constant between state 
$\epsilon_{l,k}$ in the lead $l$ and state $\epsilon_i$ in the QD $i$. 
Here we assume that the tunneling process conserves electron spins.

By using the non-equilibrium Green function method, the current for 
a steady state is given as\cite{Haug,Datta},
\begin{equation}\label{eq:cur}
  I = \sum_{\sigma}\frac{e}{h} \int dE\; \mathrm{Tr}
  \bigl( \Sigma^{L,<}_{\sigma}G^{>}_{\sigma} -
  \Sigma^{L,>}_{\sigma}G^{<}_{\sigma} \bigr),
\end{equation}
where $G^{<(>)}_{\sigma}$ is the lesser (greater) Green function of
the central region with spin $\sigma$. $\Sigma^{L,<(>)}_{\sigma}$ is
the lesser (greater) self-energy matrix of the spin $\sigma$ electron 
in the central region with the contribution from L-lead only. 
Below, we shall briefly outline the steps of calculating these 
Green functions. We shall see that the self-energies can be treated 
as parameters and their values can be chosen according to our 
considerations. 

The lesser (greater) Green function can be calculated by the
Keldysh equations,
\begin{equation}\label{eq:Keldysh}
  G^{<(>)}_{\sigma} =
  G^{r}_{\sigma}\Sigma^{<(>)}_{\sigma}G^{a}_{\sigma}, 
\end{equation}
where $G^{r(a)}_{\sigma}$ is the retarded (advanced) Green function
of the central region with spin $\sigma$. $\Sigma^{<(>)}_{\sigma}$ is
the lesser (greater) self-energy of spin $\sigma$ electron in the 
central region with the contribution from both leads. 

The matrix elements of the retarded and advanced Green functions are 
defined as
\begin{subequations}
  \label{eq:Gra}
  \begin{align}
  G^{r}_{\sigma,ij}(t-t^\p) &= -i\theta(t-t^\p)
  \aver{\acom{\anns{d}{i}(t), \cres{d}{j}(t^\p)}}, \\
  G^{a}_{\sigma,ij}(t-t^\p) &= i\theta(t^\p-t)
  \aver{\acom{\anns{d}{i}(t), \cres{d}{j}(t^\p)}}, 
  \end{align}
\end{subequations}
where $\theta(t)$ is the step function, and $\{a,b\} = ab + ba$ is the
 Fermion anti-commutator.

The equation-of-motion method is used to calculate the retarded
(advanced) Green function, $G^{r(a)}_{\sigma}$.
The time derivative of the retarded (advanced) Green function
$G^{r(a)}_{\sigma}$ is
\begin{equation}\label{eq:eomG}
i\frac{\partial}{\partial t} G^{r(a)}_{\sigma} =
  \delta(t-t^\p)I + AG^{r(a)}_{\sigma} + UG^{r(a)(2)}_{\sigma} +
  \Sigma^{r(a)}_{\sigma},
\end{equation}
where
\begin{equation}
  A = 
  \begin{pmatrix}
    \epsilon_1 & t \\
    t & \epsilon_2
  \end{pmatrix}, \qquad 
  U = 
  \begin{pmatrix}
    U_1 & 0 \\
    0 & U_2
  \end{pmatrix},
\end{equation}
and $I$ is the unity matrix. $\Sigma^{r(a)}_{\sigma}$ is the
retarded (advanced) self-energy with spin $\sigma$. The elements of
the second order Green function $G^{r(a)(2)}_{\sigma}$ are defined as
\begin{equation}
  G^{r(a)(2)}_{\sigma,ij} = \mp i\theta(\pm t \mp t^\p)
  \aver{\acom{\anns{d}{i}(t)\ann{n}{i\bar{\sigma}}(t),
      \cres{d}{j}(t^\p)}},
\end{equation}
where $\bar{\sigma}$ is the opposite spin to $\sigma$, and
$\ann{n}{i\bar{\sigma}} = \crebs{d}{i}\annbs{d}{i}$.

Usually the equation-of-motion of the Green function 
$G^{r(a)}_{\sigma}$ contains higher order correlation functions, 
for example, $G^{r(a)(2)}_{\sigma}$ in Eq.~(\ref{eq:eomG}). 
When we calculate the equations-of-motion of these higher order 
correlation functions, even higher order correlation functions appear.
In order to close those equations, we must truncated those equations 
at an appropriate order. Here, we only keep those equations containing
terms of correlation functions up to the second order.
Then, we can get the expression of $G^{r(a)}_{\sigma}$ by
solving these equations. Since the calculation process is well
established, we don't give it explicitly here. The detail calculation
can be found in Ref.~\cite{Haug}.

The retarded (advanced) Green function $G^{r(a)}_{\sigma}$ contains 
terms with the average electron numbers of the opposite spin,
$\aver{\ann{n}{i\bar{\sigma}}} \; (i = 1,2)$, in QD $i$. 
The average electron number in QD $i$ can be calculated by
\begin{equation}
  \aver{\ann{n}{i\bar{\sigma}}} = \int \frac{dE}{2\pi} \;
  \mathrm{Im} G^{<}_{\bar{\sigma},ii}(E).
\end{equation}
Thus, we need to calculate them self-consistently. After evaluating the 
average electron numbers in both QDs self-consistently, we can obtain 
all Green functions. Because we consider here a system with two QDs,
all these Green functions and self-energies are $2 \times 2$ matrices.

As shown above, we need to know self-energies to calculate those Green
functions. The elements of lesser (greater) self-energy
$\Sigma^{<(>)}_{\sigma}$ are given as
\begin{align}
  \label{eq:selfElg} \Sigma^{<(>)}_{\sigma,ij} &= \sum_{l,k}
  V_{l,k\sigma,i} V_{l,k\sigma,j}^* g^{l,<(>)}_{\sigma,k} \nonumber \\
  &= \Sigma_{\sigma, ij}^{L,<(>)} + \Sigma_{\sigma,ij}^{R,<(>)},
\end{align}
where 
\begin{subequations}
\begin{gather}
\Sigma^{L,<(>)}_{\sigma,ij} = \sum_{k} V_{L,k\sigma,i}
V_{L,k\sigma,j}^* g^{L,<(>)}_{\sigma,k}, \\
\Sigma^{R,<(>)}_{\sigma,ij} = \sum_{k} V_{R,k\sigma,i}
V_{R,k\sigma,j}^* g^{R,<(>)}_{\sigma,k},
\end{gather}
\end{subequations}
are the lesser and greater self-energies with contributions from both 
L- and R-leads, respectively. $g^{l,<(>)}_{\sigma}$ is lesser
(greater) Green function for lead $l$ with spin $\sigma$ and the
elements of these two Green functions are given as
\begin{subequations}
\begin{align}
  g^{l,<}_{\sigma,k}(t-t^\p) &=
  i\aver{\cre{c}{l,k\sigma}(t^\p)\ann{c}{l,k\sigma}(t)} \nonumber \\
  &= i f_l(\epsilon_{l,k}) e^{-i\epsilon_{l,k}(t-t^\p)}, \\
  g^{l,>}_{\sigma,k}(t-t^\p) &=
  -i\aver{\ann{c}{l,k\sigma}(t)\cre{c}{l,k\sigma}(t^\p)} \nonumber \\
  &= -i \bigl(1-f_l(\epsilon_{l,k})\bigr) e^{-i\epsilon_{l,k}(t-t^\p)},
\end{align}
\end{subequations}
where $f_l(\epsilon)=(1+e^{\beta(\epsilon - \mu_l)})^{-1}$ is the
Fermi-Dirac distribution function. $\mu_l$ is the chemical potential
in lead $l$ and $\beta=1/(k_BT)$. The time dependent operators in lead
$l$ are $\ann{c}{l,k\sigma}(t) = e^{-i\epsilon_{l,k}
  t}\ann{c}{l,k\sigma}$, $\cre{c}{l,k\sigma}(t) = e^{i\epsilon_{l,k}
  t}\cre{c}{l,k\sigma}$.

The elements of the retarded (advanced) self-energy are given as
\begin{equation}
  \label{eq:selfEra}
  \Sigma^{r(a)}_{\sigma,ij} = \sum_{l,k} V_{l,k\sigma,i}
  V_{l,k\sigma,j}^* g^{l,r(a)}_{\sigma,k},
\end{equation}
where $g^{l,r(a)}_{\sigma}$ is the retarded (advanced) Green function
of lead $l$ with spin $\sigma$ and their elements are given as
\begin{align}
   g^{l, r(a)}_{\sigma,k}(t-t^\p) &= \mp i\theta(\pm t \mp t^\p)
 \aver{\acom{\ann{c}{l,k\sigma}(t),\cre{c}{l,k\sigma}(t^\p)}}
 \nonumber \\ 
        &= \mp i\theta(\pm t \mp t^\p)e^{-i\epsilon_{l,k}(t-t^\p)}.
\end{align}

In our calculation, we need to know the Fourier transformation of the
elements of the retarded (advanced) self-energy. They are 
\begin{align}\label{eq:selfE1}
  \Sigma^{r(a)}_{\sigma,ij}(E) &= \sum_{l,k} V_{l,k\sigma,i}
  V_{l,k\sigma,j}^* g^{l,r(a)}_{\sigma,k}(E) \nonumber \\ 
          &= \Lambda^{r(a)}_{\sigma,ij}(E) \mp
          \frac{i}{2}\Gamma^{r(a)}_{\sigma,ij}(E).
\end{align}
where the real and imaginary parts, that is, the level-shift function
$\Lambda^{r(a)}_{\sigma}(E) = \Lambda^{L,r(a)}_{\sigma}(E) +
\Lambda^{R,r(a)}_{\sigma}(E)$, and the level-width function
$\Gamma^{r(a)}_{\sigma}(E) = \Gamma^{L,r(a)}_{\sigma}(E) +
\Gamma^{R,r(a)}_{\sigma}(E)$ are due to the tunneling process between 
the QDs and both L- and R-leads. 
Usually, these functions are energy-dependent. However, under the 
wide band approximation, they do not depend on electron energy. 
We shall use this wide band approximation and assume the 
level-shift and level-width functions to be independent of energy. 
In the absence of a magnetic field, they do not depend on spin 
index $\sigma$, and we shall drop it in our notations below.
Therefore, we can define those self-energies as parameters with four 
$2 \times 2$ constant matrices $\Lambda^{L(R)}$ and $\Gamma^{L(R)}$ 
instead of using the coupling constant $\{V_{l,k\sigma,i}\}$.
Then the retarded (advanced) self-energy is
\begin{equation}
\Sigma^{r(a)}_{\sigma} = \Lambda^{L} + \Lambda^{R}
          \mp \frac{i}{2}(\Gamma^{L} + \Gamma^{R})
\end{equation}
According to Eq.~(\ref{eq:selfElg}), the
lesser self-energy is given as
\begin{equation}
  \Sigma^{<}_{\sigma}(E) 
   =  i \bigl( \Gamma^{L}f_L(E) + \Gamma^{R}f_R(E) \bigr)
\end{equation}
and the greater self-energy is
\begin{equation}
  \Sigma^{>}_{\sigma}(E)
    = -i \bigl( \Gamma^{L}(1 - f_L(E)) +  \Gamma^{R} (1 -
    f_R(E))\bigr).
\end{equation}
The lesser and greater self-energies depend on energy only through the
Fermi-Dirac distribution function $f_{L(R)}(E)$. 

The Green functions can be calculated by equation-of-motion method 
described above in terms of all the self-energies. Then we can 
evaluate the $I$-$V$ curve by using Eq.~(\ref{eq:cur}).
In the next section, we shall give some numerical results of $I$-$V$
curves for different cases.

\section{Numerical results and discussion}
 \label{sec:results}

\subsection{In the absence of e-e interactions in both QDs}

First, we consider a simple case that the electron-electron (e-e) 
interactions in both QDs are absent, i.e., $U_1 = U_2 = 0$. 
The energy needed to add an extra electron in a QD due to the 
e-e interaction is $e^2/C$, where $C$ is the capacitance of the QD.
When the size of a QD are large, this energy may be very small 
compared to the level spacing. We can then ignore the e-e 
interactions in the QD. Thus, this may be used to describe a relative 
large QD. 

In order to investigate the resonance coupling effect on the $I$-$V$ 
characteristics, we assume that as shown in Fig.~\ref{fig:system} one
of two QDs (QD1) is connected to both external leads while the other
one (QD2) is only connected to the right lead. When no external bias
is applied, electronic state energy $\epsilon_1(0)$ in QD1 is set to 
be smaller than energy $\epsilon_2(0)$ in QD2. The external bias is 
applied by increasing chemical potential $\mu_L$ of the left lead and 
keeping $\mu_R$ of the right lead unchanged. We assume that the barriers 
between QD1 and left and right leads are symmetric, and voltage drops 
uniformly across the QD1. Thus, the energy level $\epsilon_1$ of QD1 
shifts to $\epsilon_1 = \epsilon_1(0) + eV/2$ under a bias $V$ bias.
In reality, because of the difference of the electro-static 
potentials of QD1 and R-lead, there is a voltage drop in QD2 and energy 
$\epsilon_2$ in QD2 should also shift. However, $\epsilon_1$ shifts 
more than $\epsilon_2$ does. When a certain bias is applied, 
$\epsilon_1$ and $\epsilon_2$ can be tuned to be aligned with each other,
or resonance coupling. In fact, other model parameters may also change. 
It is known that bias dependence of other model parameters may also 
affect $I$-$V$ characteristics\cite{sdwang}. However, for simplicity, 
we assume that the energy level $\epsilon_2$ in QD2 remains unchanged 
because we are interested in the resonance coupling effect on the 
$I$-$V$ curves in this study. This assumption may affect the positions 
of $I-V$ peaks mentioned below, it shall not change the physics studied. 
We shall come back to this point in our discussions.

Accordingly, we set the energies of two electronic states in the two
QDs at zero bias, $\epsilon_1(0) = 1.0 $, $\epsilon_2(0) = 3.5$.
Since QD1 are connected to both leads and QD2 are only connected to
the right leads, we set $\Gamma^{L}_{11}=\Gamma^{R}_{11}=0.2$,
$\Gamma^{L}_{22} = 0.0$, $\Gamma^{R}_{22} = 0.2$, $\Gamma^{L}_{12} =
\Gamma^{L}_{21}=0.0$, $\Gamma^{R}_{12}=\Gamma^{R}_{21}=0.1$. Because
the level-shift functions $\Lambda^{L(R)}$ only add a constant energy
shifting to the electronic state in each QD, their values don't change
our final results. We set them to zero, that is, $\Lambda^{L(R)}_{ij}
= 0.0$, where $i,j = 1,2$. The chemical potentials of the left lead 
$\mu_L=eV$ can vary while $\mu_R=0$. All these parameters are in 
an arbitrary unit of energy $\Gamma$ which is a system parameter.

\begin{figure}[htbp]
  \begin{center}
    \includegraphics[width=6.7cm,height=9.0cm]{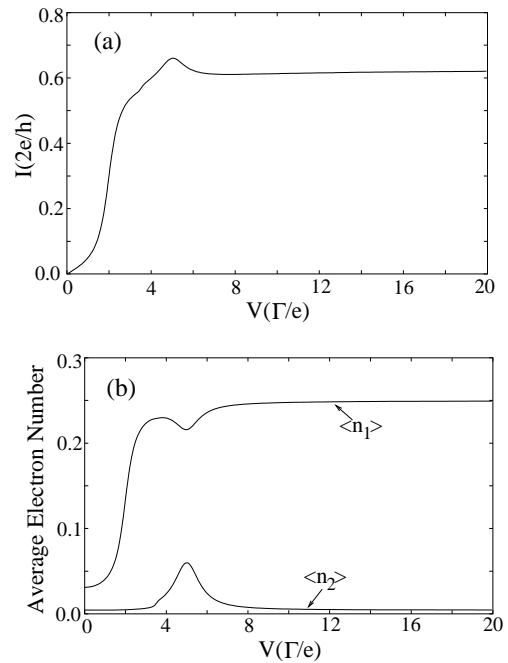}
  \end{center}
  \caption{Numerical results of $I$-$V$ and $\aver{\ann{n}{i}}$-$V$ 
    in the absence of e-e interactions. The parameters are 
    $\epsilon_1(0) = 1.0$, $\epsilon_2(0) = 3.5$, $\Gamma^{L}_{11}
    = \Gamma^{R}_{11} = 0.2$, $\Gamma^{L}_{22} = 0.0$,
    $\Gamma^{R}_{22} = 0.2$, $\Gamma^{L}_{12} = \Gamma^{L}_{21} =
    0.0$, $\Gamma^{R}_{12} = \Gamma^{R}_{21} = 0.1$,
    $\Lambda^{L(R)}_{ij} = 0.0$ ($i,j=1,2$), $\mu_L=eV$ and $\mu_R=0$,
    where all these parameters are in unit of $\Gamma$. The
    temperature $T=0.01$ in unit of $\Gamma/k_B$. (a) Current $I$ vs
    applied bias $V$. The current $I$ has a peak at $V=5.0$. (b)
    Average electron numbers in both QDs, $\aver{\ann{n}{i}}$ vs applied
    bias $V$. $\aver{\ann{n}{1}}$ has a valley while
    $\aver{\ann{n}{2}}$ has a peak at $V=5.0$.}
  \label{fig:ivq-nonU}
\end{figure}

We first calculate $I$-$V$ curves at a very low temperature, 
$T = 0.01$ in the unit of $\Gamma/k_B$, so that we can neglect 
the thermal broadening of energy levels in QDs. 
Fig.~\ref{fig:ivq-nonU}(a) shows the calculated $I$-$V$ curve and
Fig.~\ref{fig:ivq-nonU}(b) shows the numerical result of the bias 
dependence of the average electron numbers $\aver{\ann{n}{i}}$ in
QD $i$.
At low and high bias, the two electronic states are far apart.
Since QD2 is connected to the right lead only, electrons can only 
tunnel into it through QD1. In the absence of any inelastic processes, 
the chance for an electron tunneling from QD1 to QD2 is very 
small when the two electronic states, $\epsilon_1$ and $\epsilon_2$, 
are far apart. Therefore, the average electron number in QD2 is almost 
zero. Its effect on the tunneling process is negligible, and the $I$-$V$ 
characteristics should be very similar to that of a single QD system. 
This is indeed what we can see from Fig.~\ref{fig:ivq-nonU}(a). 
As shown in Fig.~\ref{fig:ivq-nonU}(a), there is a peak at $V=5.0$,
where two electronic states $\epsilon_1$ and $\epsilon_2$ are aligned
with each other. Correspondingly, the average electron number in QD1, 
$\aver{\ann{n}{1}}$ has a valley at this bias as shown in 
Fig.~\ref{fig:ivq-nonU}(b). At the same time, there is a dramatical 
increase in the average electron number in QD2. This is not surprising
at all. The chance for an electron tunneling from one electronic state 
in QD1 to another state in QD2 increases as the energies of the two 
states approach each other. It become maximum when they are precisely 
aligned. We shall call it resonance coupling of the two QDs when this 
happens. At resonance coupling, a new tunneling channel through QD2 opens. 
In another words, the effective tunneling rate of an electron out of the 
QD dimer increases. This causes the peak observed in 
Fig.~\ref{fig:ivq-nonU}(a), leading to a NDR.

This can be understood from the following argument. 
As the electron tunneling probability from QD1 to QD2 increases,
the effective tunneling rate out of the electronic state $\epsilon_1$
increases and the effective tunneling rate into the electronic state
$\epsilon_2$ increases as well.  Therefore, $\aver{\ann{n}{1}}$
decreases and $\aver{\ann{n}{2}}$ increases because more electrons
tunnel from QD1 to QD2. For the sequential tunneling, 
the resonant tunneling current and the average electron number in the
resonant level at zero temperature are given as\cite{Datta}
\begin{subequations}\label{eq:qI}
\begin{align}
\aver{n} &\propto \frac{\Gamma^\mathrm{in}}{\Gamma^\mathrm{in} +
\Gamma^\mathrm{out}} \\
I &\propto
\frac{\Gamma^\mathrm{in}\Gamma^\mathrm{out}}{\Gamma^\mathrm{in} +
\Gamma^{\mathrm{out}}}
\end{align}
\end{subequations}
where $\Gamma^{\mathrm{in}(\mathrm{out})}$ is the tunneling rate 
of an electron into (out of) the resonant level. Near the resonance 
coupling, for QD1, the tunneling rate $\Gamma^\mathrm{in}_1$ from the 
left lead into the dot remains almost unchanged and the tunneling rate 
$\Gamma^\mathrm{out}_1$ out of the dot increases because of a new 
tunneling channel through QD2. For QD2, the tunneling rate 
$\Gamma^\mathrm{in}_2$ from QD1 into QD2 increases while the tunneling 
rate $\Gamma^\mathrm{out}_2$ out of QD2 into the right lead remains 
unchanged. Therefore, the overall tunneling rates for the QD dimer is 
as follows. The tunneling rate $\Gamma^\mathrm{in}$ from left lead 
into the QD dimer is almost unchanged, while the effective tunneling 
rate $\Gamma^\mathrm{out}$ out of the QD dimer increases. Therefore, 
the total current increases and $\aver{\ann{n}{1}}$ decreases 
while $\aver{\ann{n}{2}}$ increases according to Eq.~(\ref{eq:qI}).
In turn, it generates a $I$-$V$ peak and a NDR. 
\begin{figure}[htbp]
  \begin{center}
    \includegraphics[width=6.0cm,height=8.2cm]{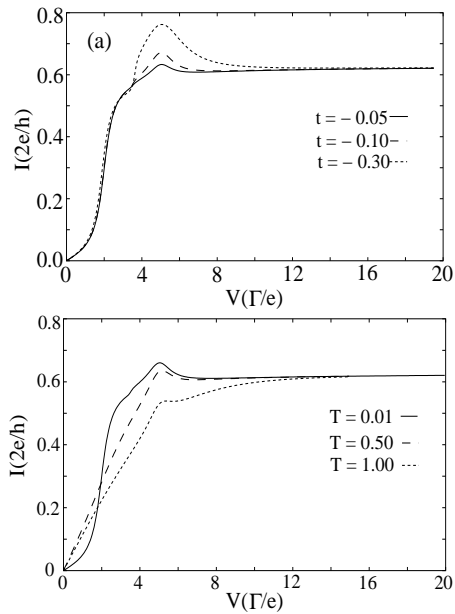}
    \end{center} \caption{Numerical results of $I$-$V$ curves at
      different hopping energies and temperatures. All other parameters
      are the same as those in Fig.~\ref{fig:ivq-nonU}. (a) $I$-$V$ at
      different hopping energies, $t=-0.05$ (solid line), $t=-0.10$
      (dashed line), and $t=-0.30$ (dotted line). The peak height and
      width increase as the hopping energy $t$. (b) $I$-$V$ at
      different temperatures, $T= 0.01$ (solid line), $T=0.50$ (dashed
      line), and $T=1.00$ (dotted line). The peak is
      smeared at high temperatures.}
    \label{fig:ivhT-nonU}
\end{figure}

We have seen that resonance coupling between two dots can lead to a 
peak in the $I$-$V$ curve. Thus, we should expect the height and width of 
this peak to depend on the interdot coupling strength. Also, the 
thermal energy can introduce inelastic tunneling processes which can 
wash out the resonance effect. One should also expect that the peak is 
also sensitive to the temperature. 
The $I$-$V$ curves at different interdot coupling strengths between
two QDs, and at different temperatures are shown in
Fig.~\ref{fig:ivhT-nonU}(a) and (b), respectively. As shown in 
Fig.~\ref{fig:ivhT-nonU}, the peak height and width increase as the
coupling between two QDs increases. Because the chance of an electron 
tunneling from QD1 to QD2 increases with the interdot coupling, 
more electrons can tunnel out of the dimer through QD2, and the 
effective tunneling rate out of the QD dimer increases. Therefore
the peak height and width increase as the interdot coupling increases.
We can also see the thermal smearing of the NDR. 
The peak disappears gradually with the increase of temperature.


\subsection{The presence of e-e interactions in QDs}

\begin{figure}[htbp]
  \begin{center}
    \includegraphics[width=6.7cm,height=9.0cm]{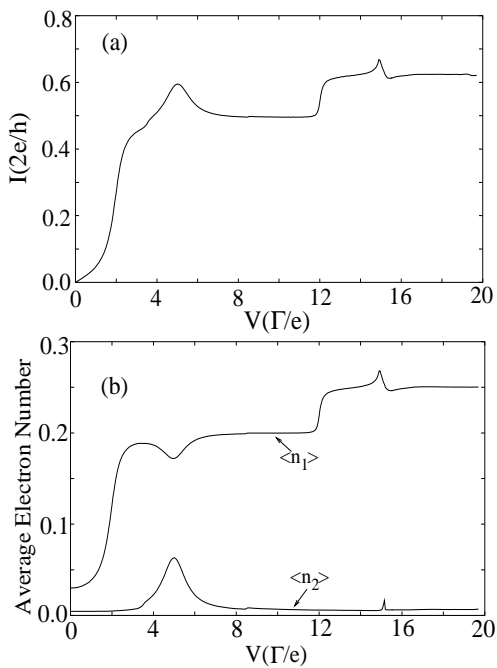}
  \end{center}
  \caption{Numerical results of $I$-$V$ and $\aver{\ann{n}{i}}$-$V$ in
    the case in presence of e-e interactions. The e-e interactions in
    two QDs are $U_1 = U_2 = 5.0$ and other parameters are the same as
    those in Fig.~\ref{fig:ivq-nonU}. (a) Current $I$ vs applied bias
    $V$. The current $I$ has two peaks. one is at $V=5.0$ and the
    other is at $V=15.0$. (b) Average electron numbers in both QDs,
    $\aver{\ann{n}{i}}$ vs applied bias $V$. $\aver{\ann{n}{1}}$ has a
    valley at $V=5.0$ and a peak at $V=15.0$. $\aver{\ann{n}{2}}$ has
    two peaks. One is at $V=5.0$ and the other is at $V=15.0$.}
  \label{fig:ivq-U} 
\end{figure}

In order to see whether the $I$-$V$ peaks, thus the NDR, due to the 
resonance coupling of two QDs will survive when the e-e interaction is
present, 
we repeat the above calculation by including non-zero $U_1$ and $U_2$.
We set the Coulomb interactions $U_1 =U_2 =  5.0$ while keeping the 
other parameters the same as those in Fig.~\ref{fig:ivq-nonU}. 
The calculated results of $I$-$V$ curve and the bias dependence of
the average electron numbers $\aver{\ann{n}{i}}$ in QD $i$ are
shown in
Fig.~\ref{fig:ivq-U}(a) and (b), respectively. As one will expect,
the peak at $V=5.0$ is still there with the similar features as those 
in the absence of the e-e interaction. The reason for the occurrence 
of this peak should be the same as that in the previous section.

However, there is an extra peak at $V=15.0$. This new peak can be 
attributed to the e-e interactions in QDs. In order to understand the
origin of this new peak, let us examine the possible electronic 
states in QD $i$. In the case that no electron in 
the QD, an electron of energy $\epsilon_i$ can move in freely into 
it. However, only the electron with energy $\epsilon_i + U_i$ can 
jump into the QD when there is already one electron in because of 
the e-e interaction. Therefore, the electronic states of the QD 
has effectively two distinct energy values, $\epsilon_{i}$ and 
$\epsilon_{i} + U_{i}$. We have already identify the first peak 
as the resonance coupling due to the alignment $\epsilon_1$
and $\epsilon_2$. It is natural to expect another peak to appear 
when $\epsilon_1$ is aligned with $\epsilon_2 + U_2$. In our case, 
it corresponds to $V=15.0$, exactly what we see in
Fig.~\ref{fig:ivq-U}(a). In comparison with the first peak, 
the second peak is smaller and narrower. This is due to the different 
properties of the two electronic states $\epsilon_2$ and $\epsilon_2+U_2$. 
Unlike $\epsilon_2$, $\epsilon_2+U_2$ doesn't exist unless one electron 
has already been in QD2. However, when $\epsilon_1$ and $\epsilon_2 + 
U_2$ are aligned, $\epsilon_2$ and $\epsilon_1$ are far apart. 
Therefore, the probability of an electron hopping from $\epsilon_1$ to 
$\epsilon_2$ is small. Consequently, the average electron number in  
$\epsilon_2$ is very small. Thus, the existing probability  
of $\epsilon_2 + U_2$ state is very small too. This is probably the 
reason why the second peak is smaller and narrower.

Unlike the first resonance-coupling peak where $\aver{\ann{n}{2}}$
is maximum and $\aver{\ann{n}{1}}$ is minimum, both $\aver{\ann{n}{1}}$ 
and $\aver{\ann{n}{2}}$ are maximum at the second peak. 
$\aver{\ann{n}{2}}$ becomes maximum at the second resonance-coupling 
peak because the effective tunneling rate into QD2 from QD1 increases 
as $\epsilon_1$ approach $\epsilon_2+U_2$. 
The maximum of $\aver{\ann{n}{1}}$ is because both
tunneling rate into and out of QD1 increase. Around $V=15.0$, the
effective tunneling rate out of QD1 increases because more electrons
can tunnel from QD1 to QD2. On the other hand, the effective tunneling
rate into QD1 increases too. Two electrons can now be in QD1 at the 
same time, occupying $\epsilon_1$ and $\epsilon_1 + U_1$ states,
respectively. After the electron with energy $\epsilon_1$ tunnels out
of QD1, the remaining electron has energy $\epsilon_1$ instead of
its original energy $\epsilon_1 + U_1$. Then a new electron with
energy $\epsilon_1 + U_1$ can tunnel from the left lead into QD1. As
$\epsilon_1$ and $\epsilon_2 + U_2$ are aligned, electrons in
$\epsilon_1$ tunnel out of QD1 faster.  Therefore, electrons can
tunnel into QD1 from the left lead faster and the effective tunneling
rate into QD1 increases. So when $\epsilon_1$ and $\epsilon_2 + U_2$
are aligned, both the effective tunneling rate into QD1
$\Gamma^\mathrm{in}_1$ and the effective tunneling rate out of QD1
$\Gamma^\mathrm{out}_1$ increase. According to Eq.~(\ref{eq:qI}) the
average electron number in $\epsilon_1$ may increase if
$\Gamma^\mathrm{in}_1$ increases faster than $\Gamma^\mathrm{out}_1$.
This explains the possibility that $\aver{\ann{n}{1}}$ reaches maximum 
value at the second peak. 

In the above calculations, we have not considered the case that
$\epsilon_1+U_1$ is aligned with $\epsilon_2+U_2$. This is because the 
bias is too small to allow electrons tunneling through $\epsilon_1 + U_1$, 
when those two electronic states are aligned. Moreover, if we set the same
e-e interactions in two QDs, $\epsilon_1 + U_1$ and $\epsilon_2 + U_2$
are aligned at the same time when $\epsilon_1$ and $\epsilon_2$ are
aligned. These two effects may not be distinguished easily. In order
to
avoid this possible confusion, we consider different e-e interactions 
in two QDs such that $\epsilon_1$ and $\epsilon_2$ are not aligned 
with each other when $\epsilon_1+U_1$ and $\epsilon_2+U_2$ do. 
We also choose a set of parameters in such a way that electrons can
tunnel through $\epsilon_1 + U_1$ when it is aligned with 
$\epsilon_2 + U_2$.
\begin{figure}[htbp]
  \begin{center} 
    \includegraphics[width=6.0cm,height=8.5cm]{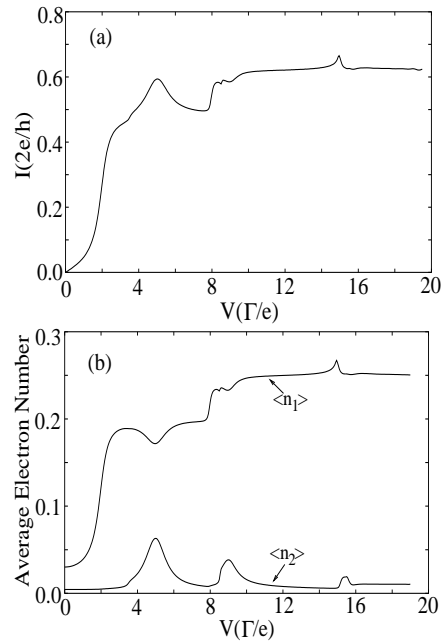}
  \end{center} 
   \caption{Numerical results of $I$-$V$ and $\aver{\ann{n}{i}}$-$V$
     in the case of asymmetric e-e interactions of two QDs. The e-e
     interactions in two QDs are $U_1=3.0$ and $U_2=5.0$. All other
     parameters are the same as those in Fig.~\ref{fig:ivq-nonU}. (a)
     Current $I$ vs applied bias $V$. The current $I$ has four
     peaks. one is at $V=5.0$, one at $V=15.0$, and other two small
     peaks at $V=8.3$ and $V=8.6$, respectively. (b) Average electron
     numbers in both QDs, $\aver{\ann{n}{i}}$ vs applied bias
     $V$. $\aver{\ann{n}{1}}$ has a valley at $V=5.0$ and three
     peaks. The first is at $V=8.3$, the second at $V=8.6$, and the
     third at $V=15.0$. $\aver{\ann{n}{2}}$ has three peaks. One is at
     $V=5.0$, one at $V=8.6$, and the last at $V = 15.0$.}
   \label{fig:ivq-U1}
\end{figure}

Fig.~\ref{fig:ivq-U1}(a) shows $I$-$V$ curve with $U_1=3.0$, $U_2 =
5.0$, and all other parameters are the same as those in
Fig.~\ref{fig:ivq-nonU}. Besides two peaks at $V=5.0$ and $V=15.0$,
there are two additional very small peaks. One is at $V = 8.3$ and 
the other at $V = 8.6$. Similar to the two old resonance-coupling 
peaks, these two new small peaks are due to the alignment of 
$\epsilon_1 + U_1$ and $\epsilon_2 + U_2$. This resonance-coupling 
yields two peaks instead of one because of the competition of 
two opposite effects that do not exist in the previous case. 
As we explain above, the existing probability of electronic state 
$\epsilon_2 + U_2$ decrease as $\epsilon_1$ and $\epsilon_2$ are far 
apart. However, when $\epsilon_1 + U_1$ and $\epsilon_2 + U_2$ become 
closer, the electron tunneling probability from $\epsilon_1 + U_1$ to
$\epsilon_2 + U_2$ increases. The competition of these two effects
leads to those two peaks in the $I$-$V$ curve. 

As shown in Fig~\ref{fig:ivq-U1}(b), $\aver{\ann{n}{1}}$ has one 
valley at $V=5.0$ and three peaks. They are located at $V=8.3$, 
$8.5$, and $15.0$, respectively. $\aver{\ann{n}{2}}$ has three peaks 
at $V = 5.0$, $8.5$, and $15.0$, respectively. The behavior of 
$\aver{\ann{n}{i}}$, $i=1,2$, at $V=5.0$ and $15.0$ has 
been already explained before. The reason of the occurrence of the
other two peaks of $\aver{\ann{n}{1}}$ is due to the competition of 
two opposite effects mentioned above. When $\epsilon_1 + U_1$ and 
$\epsilon_2 + U_2$ are aligned, there are those two opposite effects. 
One decreases the
average electron number in QD2 while the other increases it. But the
increase of the average electron number is larger than the decrease of
it, therefore we can only see one peak of $\aver{\ann{n}{2}}$.

\begin{figure}[htbp]
  \begin{center}
    \includegraphics[width=6.2cm,height=8.5cm]{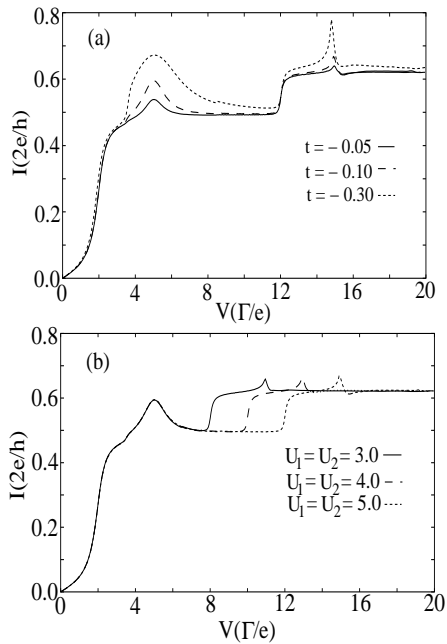}
  \end{center}
  \caption{Numerical results of $I$-$V$ curves at different coupling
    strengths and e-e interactions. All other parameters are the same as
    those in Fig.~\ref{fig:ivq-nonU}. (a) $I$-$V$ at different hopping
    energies, $t=-0.05$ (solid line), $t=-0.10$ (dashed line), and
    $t=-0.30$ (dotted line). The e-e interactions are
    $U_1=U_2=5.0$. (b) $I$-$V$ at different e-e interactions,
    $U_1=U2=3.0$ (solid line), $U_1=U_2=5.0$ (dashed line), and
    $U_1=U_2=1.00$ (dotted line).}
  \label{fig:ivhU-U}
\end{figure}

\begin{figure}[htbp]
  \begin{center}
    \includegraphics[width=5.8cm,height=4.5cm]{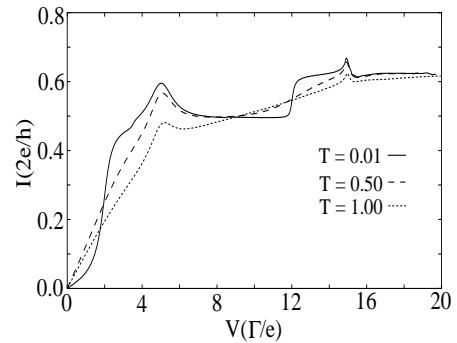}
  \end{center}
\caption{Numerical results of $I$-$V$ curves at different 
temperatures, $T= 0.01$ (solid line), $T=0.50$ (dashed line), and
$T=1.00$ (dotted line). All other parameters are the same as those 
in Fig.~\ref{fig:ivq-nonU}. Peaks are smeared at high temperatures.} 
  \label{fig:ivT-U}
\end{figure}

The $I$-$V$ curves of different coupling strengths and e-e interactions 
are shown in Fig.~\ref{fig:ivhU-U}(a) and (b), respectively.
As one will expect, the heights and widths of both peaks increase
as the coupling strength between two QDs increases.
From Fig.~\ref{fig:ivhU-U}(b), one can see 
that the e-e interactions have no effect on the peak at low bias 
because the peak at low bias is due to the alignment of $\epsilon_1$ 
and $\epsilon_2$. A strong e-e interaction shifts the position of the 
peak at high bias, but it does not change its height and width. 
The peak at high bias is due to the alignment of $\epsilon_1$ and
$\epsilon_2 + U_2$. As we have explained before, the height
and width of this peak depend on the existing probability of
$\epsilon_2 + U_2$ state. When $\epsilon_1$ and $\epsilon_2 + U_2$ are
aligned, $\epsilon_1$ and $\epsilon_2$ are far apart. Thus the
probability of an electron hopping from $\epsilon_1$ to $\epsilon_2$
is too small to be sensitive to the change of the e-e
interactions. Consequently, the existing probability of $\epsilon_2 +
U_2$ state is independent of the e-e interactions. Therefore, the
strong e-e interactions have little effect on the height and width of
this peak.
The $I$-$V$ curves at different temperatures are shown in
Fig.~\ref{fig:ivT-U}. Peaks disappear at hight temperature due to the
thermal smearing effects.

\begin{figure}[htbp]
  \begin{center}
    \includegraphics[width=4.5cm,height=5.5cm]{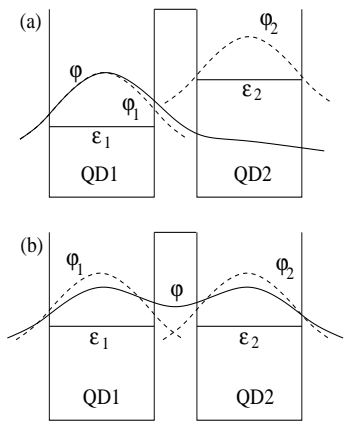}
  \end{center}
\caption{Schematic diagrams of the wave functions of an electron in (a)
  off-resonance coupling and (b) resonance coupling cases. $\varphi_1$
  and $\varphi_2$ are eigenfunctions with energy $\varepsilon_1$ and
  $\varepsilon_2$ in two QDs without interdot coupling. $\varphi$ is
  the wave function (bonding state) in the presence of the coupling.
  (a) In off-resonance coupling case, the overlap of $\varphi_1$ and
  $\varphi_2$ in two QDs is negligible and an electron in the bonding
  state is largely localized in QD1. (b) In resonance coupling case,
  the overlap of $\varphi_1$ and $\varphi_2$ are very large and
  electron in the bonding state has approximately the same probability
  of being found in either one of the two QDs.}
  \label{fig:wavefunction}
\end{figure}

What is seen is a new mechanism for the NDR in a system with more than
one QDs. The NDR is a very important phenomenon and has many
important applications. Devices with the NDR have been widely used to
make amplifier and oscillators in a very wide frequency
range\cite{sze}. In superlattice, it is known that a NDR leads to
many interesting phenomena, such as current-voltage oscillation on the
sequential resonant tunneling plateau, current self-oscillation, and
chaos\cite{xrw}. There are many mechanisms for NDR, such as Gunn
effect and resonant tunneling in superlattices. The NDR found here is
due to the resonance coupling between two QDs. As shown in
Fig.~\ref{fig:wavefunction}, the wave function of the bonding state is
largely localized in QD1 in off-resonance case. Electrons can only
tunnel from the source to the drain through QD1.
In resonance coupling case, electrons in the
bonding state can be in both QDs, and electrons can also tunnel from
the source to the drain through QD2. A new tunneling channel is open
and a peak can be observed in the $I$-$V$ curve. This leads to the
NDR.
Furthermore, the properties of the NDR depend on the properties of
both QDs. For example, the widths and heights of peaks vary as the
dot-dot coupling. And the positions of peaks depend on the energy
spectra of both QDs. The variations of the energy spectra of the QD
under the probing tip or the adjacent QD, or both QDs may change the
positions of peaks in $I$-$V$ curves. Thus, the $I$-$V$
characteristics shall have different behavior when the tip is above
QD2 rather than QD1. Comparing this difference will allow one to
distinguish a NDR due to the mechanism proposed here from the others.
A better understanding of this resonance coupling effect may
also enhance the STM as a powerful probe not only for a regular
surface, but also for a cluster structure.

We show that an $I$-$V$ peak, thus NDR, appears at the resonance
coupling between two QDs where their energy levels are aligned. For
simplicity, we consider essentially only one electronic state in each
QD. The mechanism should survive in the case that there are more than
one electronic state in each QD. This effect should be pronounced when
the sizes of QDs are small such that the discrete natural of
electronic state are clear. We assume also that the bias shifts only
the energy levels in the QD directly connected to two leads and has no
effect on other parameters. In reality, the bias not only shifts the
energy levels of both QDs but also changes the tunneling
rates\cite{sdwang}. But in our quantum dot configuration, which may
arise in a STM experiments, the energy levels in QD1 shall be affected
more than those in QD2 by an external bias. Thus, the resonance
coupling will occur under certain bias. We use this simplified
assumption in order to unambiguously identify the observed $I$-$V$
peaks. Since the physics is due to the resonance coupling which indeed
should occur in real experiments, we believe this new mechanism is
very robust, and does not depend on our assumption. Of course, the
peak positions, which are not our main concern, depend surely on the
assumption. In fact, the NDR due to the resonance coupling of two QDs
may have been observed already in a recent experiment\cite{gwang}.


\section{Summary}
\label{sec:summary}

In summary, the dot-dot coupling effect on the tunneling current under 
a bias is investigated. We considered a QD dimer connected to two 
metallic leads in such a way that one QD in the dimer is connected to 
the both leads while the other is only directly connected to one lead. 
We show numerically that peaks, thus NDRs, appear in the $I$-$V$ curves 
at resonance coupling when two electronic states in two QDs are aligned. 
This NDR exists both in the absence and in the presence of 
e-e interactions. We study how the peaks in the $I$-$V$ characteristics
change with interdot coupling, e-e interaction, and the temperature.
We show that the heights and widths of those peaks increase with 
the interdot coupling of the QD dimer. But they are almost unaffected by 
the e-e interactions in both QDs when the e-e interactions are strong.
We show that those peaks are smeared at high temperatures due to 
the thermal smearing effects.


\begin{acknowledgments}
We would like to acknowledge the support of the Research Grant
Council of HKSAR, China.
\end{acknowledgments}


\end{document}